\documentclass[utf8]{FrontiersinHarvard} 

\usepackage{url,hyperref,lineno,microtype,subcaption}
\usepackage[onehalfspacing]{setspace}
\usepackage{booktabs}

\def\keyFont{\fontsize{8}{11}\helveticabold }
\def\firstAuthorLast{Hu {et~al.}} 

\def\Authors{Chuanbo Hu\,$^{1}$,Jacob Thrasher$^{4}$,Wenqi Li\,$^{1}$, Mindi Ruan$^{4}$, Xiangxu Yu\,$^{3}$, Lynn K Paul\,$^{2}$, Shuo Wang\,$^{3}$, and Xin Li\,$^{1,*}$}
\def\Address{$^{1}$Department of Computer Science, University at Albany, Albany, NY 12222, USA \\
$^{2}$Humanities and Social Sciences, California Institute of Technology, Pasadena, CA 91125, USA \\
$^{3}$Department of Radiology, Washington University in St. Louis, St. Louis, MO 63110, USA \\
$^{4}$Lane Department of Computer Science and Electrical Engineering, West Virginia University, Morgantown, WV, USA }

\def\corrAuthor{Corresponding Author}

\def\corrEmail{xli48@albany.edu}

\begin{document}

\onecolumn
\firstpage{1}

\title[Running Title]{Exploring Speech Pattern Disorders in Autism using Machine Learning} 

\author[\firstAuthorLast ]{\Authors} 
\address{} 
\correspondance{} 

\extraAuth{}

\maketitle

\begin{abstract}

Diagnosing autism spectrum disorder (ASD) by identifying abnormal speech patterns from examiner-patient dialogues presents significant challenges due to the subtle and diverse manifestations of speech-related symptoms in affected individuals. This study presents a comprehensive approach to identify distinctive speech patterns through the analysis of examiner-patient dialogues. Utilizing a dataset of recorded dialogues, we extracted 40 speech-related features, categorized into frequency, zero-crossing rate, energy, spectral characteristics, Mel Frequency Cepstral Coefficients (MFCCs), and balance. These features encompass various aspects of speech such as intonation, volume, rhythm, and speech rate, reflecting the complex nature of communicative behaviors in ASD. We employed machine learning for both classification and regression tasks to analyze these speech features. The classification model aimed to differentiate between ASD and non-ASD cases, achieving an accuracy of 87.75\%. Regression models were developed to predict speech pattern related variables and a composite score from all variables, facilitating a deeper understanding of the speech dynamics associated with ASD. The effectiveness of machine learning in interpreting intricate speech patterns and the high classification accuracy underscore the potential of computational methods in supporting the diagnostic processes for ASD. This approach not only aids in early detection but also contributes to personalized treatment planning by providing insights into the speech and communication profiles of individuals with ASD.

\section{}


\tiny
 \keyFont{ \section{Keywords:} speech pattern, ASD, machine learning, ADOS, audio, medical dialogues} 
\end{abstract}

\section{Introduction}

Autism spectrum disorder (ASD) is a developmental condition characterized by significant challenges in social interaction, communication, and behavioral patterns \cite{leekam2011restricted,lord2018autism,lord2020autism}. The condition affects approximately 1 in 36 children and 1 in 45 adults in the United States, highlighting a substantial public health concern \cite{maenner2020prevalence,dietz2020national}. Despite its high prevalence, diagnosing ASD remains complex and fraught with difficulties. The diagnosis process is highly subjective, heavily reliant on behavioral observations and the clinical judgment of specialists, which can lead to disparities in the accuracy and timeliness of diagnosis across different regions and demographics.

Traditionally, ASD diagnoses are conducted through clinical interviews and behavioral observations, according to standardized diagnostic tools such as the Autism Diagnostic Observation Schedule (ADOS) \cite{lord1999ados}. These methods, while effective, rely heavily on the clinician's expertise and can be influenced by the subjective judgments of the observer, leading to potential discrepancies in diagnosis. Moreover, the current procedures are not only labor-intensive but also require the presence of trained specialists, who are often in short supply, particularly in under-resourced areas.


Early and accurate diagnosis of ASD can greatly enhance intervention outcomes, but the current reliance on traditional methods presents significant barriers. Speech, as a primary mode of communication, carries rich markers, such as prosodic features (intonation, volume, rhythm, and rate), articulation, and phonetic characteristics, which are often altered in individuals with ASD \cite{mody2013speech,pickles2009loss,vogindroukas2022language}. Analyzing speech markers with high precision can lead to more objective, reliable, and scalable diagnostic processes, providing a non-invasive, cost-effective, and rapid method of detection that could supplement and streamline traditional diagnostic processes.

Speech patterns in ASD are distinctive; they often include atypical intonation, lack of rhythm, and peculiarities in sound production and response \cite{nakai2014speech,sharda2010sounds,mayer2016mapping}. These patterns are not only prevalent but also among the most consistently observed traits across the spectrum. However, the subtlety and complexity of these speech patterns pose significant challenges for traditional analysis methods, which may lack the sensitivity or scalability to handle nuanced variations effectively. The quantifiable nature of speech parameters—such as pitch, rate, and consistency—makes them ideal for computational analysis.

Recently, Machine Learning (ML) has been employed to analyze various behavioral markers in ASD \cite{wang2015atypical,ruan2021deep,ruan2023video,zhang2022discriminative}. For instance, the early identification of autism can leverage ML-based digital behavioral phenotyping techniques \cite{perochon2023early}. Using deep learning, it's possible to recognize behaviors such as hand flapping from informal home videos, which assists in diagnosing autism \cite{lakkapragada2022classification}. Studies using ML have shown that autistic toddlers display unique patterns in head movement dynamics while engaged with audiovisual stimuli \cite{krishnappa2023complexity}. Additionally, ML has been utilized to evaluate motor imitation skills and the ability to learn by imitation \cite{lidstone2021automated,tunccgencc2021computerized,zampella2021computational}, as well as to identify unusual attention patterns in toddlers diagnosed with autism \cite{campbell2019computer}. Natural language processing techniques have been employed on electronic health records to create a detailed phenotype ontology for ASD \cite{zhao2022development}. Moreover, machine learning has been instrumental in defining the diagnostic boundaries between ASD and non-ASD cases \cite{tuncc2021diagnostic}. Furthermore, in response to the disruptions caused by the COVID-19 pandemic, a tele-assessment tool for ASD, named TELE-ASD-PEDS \cite{wagner2021use}, was developed using machine learning applied to a comprehensive clinical database encompassing several hundred children \cite{adiani2019usability}.

This research aims to harness the power of machine learning to overcome these challenges by:
\begin{itemize}
    \item \textbf{Extracting a Wide Range of Speech Features}: Utilizing advanced signal processing techniques to extract 40 different speech features covering various aspects such as intonation, volume, rhythm, rate, pause and duration from examiner-patient dialogues.
    \item \textbf{Employing Machine Learning for Analysis}: Applying machine learning algorithms to classify dialogues into ASD and non-ASD cases and to perform regression analyses on the extracted features to predict prosody-related variables and a comprehensive ASD-related score.

    \item \textbf{Improving Diagnostic Accuracy and Efficiency}: Enhancing the objectivity, accuracy, and efficiency of ASD diagnosis by reducing the subjectivity associated with traditional diagnostic methods and by providing a scalable solution to analyzing extensive speech data.
\end{itemize}


By integrating machine learning with speech analysis, this study provides a methodological advancement in the ASD diagnosis. The use of comprehensive speech feature sets and sophisticated analytic techniques represents a significant improvement over existing diagnostic practices, promising not only higher diagnostic accuracy but also earlier detection capabilities. Ultimately, this research aims to contribute to the personalized treatment and management of ASD, facilitating better outcomes for individuals affected by the disorder.

\section{Methods}

\subsection{Caltech Audio Dataset}
To implement the proposed objectives, we conducted our ASD evaluations using the Caltech audio dataset administered through the Autism Diagnostic Observation Schedule (ADOS) \cite{lord1999ados,american2013diagnostic}. This involves structured, yet naturally flowing conversations between the interviewers and the adult participants, effectively highlighting the complex behavioral patterns associated with ASD. 

\subsubsection{Autism Diagnostic Observation Schedule (ADOS)}

The Autism Diagnostic Observation Schedule, Second Edition (ADOS-2) \cite{lord1999ados} serves as an updated version of the original ADOS, which is a standardized instrument for diagnosing ASD. ADOS-2 facilitates the assessment of communication, social interaction, and restrictive and repetitive behaviors through a mixture of structured and semi-structured tasks. 
These tasks are designed to simulate social interactions between the examiner and the examinee. Specifically, Module 4 of ADOS-2, intended for verbally fluent adolescents and adults, is utilized for individuals typically aged 16 years and older. 
This module contrasts with others that are tailored based on the age and verbal skills of subjects, ranging from non-verbal children to verbally capable younger children. A2 score particularly assesses abnormalities in speech patterns such as intonation, volume, rate, and rhythm as shown in Table \ref{tab:a2}.


\begin{table}[h!]
\centering
\small
\begin{tabular}{|c|p{15cm}|}
\hline
\textbf{Score} & \textbf{Description} \\ \hline
0 & Appropriately varying intonation, reasonable volume, and normal rate of speech, with regular rhythm coordinated with breathing. \\ \hline
1 & Little variation in pitch and tone; rather flat or exaggerated intonation, but not obviously peculiar, OR slightly unusual volume, AND/OR speech that tends to be somewhat unusually slow, fast, or jerky. \\ \hline
2 & Speech that is clearly abnormal for ANY of the following reasons: slow and halting; inappropriately rapid; jerky and irregular in rhythm (other than ordinary stutter/stammer), such that there is some interference with intelligibility; odd intonation or inappropriate pitch and stress; markedly flat and toneless ("mechanical"); consistently abnormal volume. \\ \hline
7 & Stutter or stammer or other fluency disorder (if odd intonation is also present, code 1 or 2 accordingly). \\ \hline
\end{tabular}
\caption{Speech Abnormalities Associated With Autism (Intonation/Volume/Rhythm/Rate).}
\label{tab:a2}
\end{table}

\subsubsection{ADOS Interview Participants and Video Acquisition}

ADOS session was conducted sequentially across 15 different sections, involving various standardized questions and instructions aimed at eliciting a range of responses. In these sessions, the interviewer typically asks a series of questions, to which the participants respond. The interactions are structured around 15 scenario tasks designed to provoke communicative and social responses indicative of ASD. These scenarios comprehensively cover a diverse array of social interactions and communicative behaviors.

\subsubsection{ADOS Interview Audios}

The ADOS interviews, which focus on semi-structured assessments of communication, social interaction, and imaginative use of materials, are pivotal for diagnosing suspected cases of autism. In this study, interviews with all participants suspected of having ASD were recorded. The video recordings were meticulously analyzed and scored by clinical psychologists certified in ADOS administration. These scores are crucial as they form the foundation for training our machine learning models to recognize ASD-related patterns. A total of 44 subjects participated in this study, providing a robust dataset for analysis.

\subsection{Feature Extraction for Identification of Autism Speech Disorder}

Feature extraction is a critical step in analyzing speech data, especially when dealing with complex disorders like ASD. This process involves quantifying various aspects of speech that may indicate the presence of ASD traits. We extracted a comprehensive set of features from the recorded dialogues, categorized into different groups based on their characteristics and relevance to ASD. The prosodic speech features including number of syllables, number of pauses, rate of speech, articulation rate, speaking duration, original duration, balance, and frequency were meticulously extracted using the tool 'myprosody' \cite{Shahab_myprosody}. This tool integrates various speech feature extraction methodologies, providing a comprehensive analysis of prosody. Additionally, other audio features such as MFCCs, spectrograms, and chromagrams were extracted using the tool 'pyAudioAnalysis' \cite{giannakopoulos2015pyaudioanalysis}, further enriching our dataset with diverse audio representations crucial for analyzing speech patterns associated with ASD.These features are described below and summarized in Table \ref{table:speech_features}.


\begin{table}[ht]
\centering
\small
\begin{tabular}{|p{0.4cm}|p{1.36cm}|p{3.1cm}|p{10.5cm}|p{0.2cm}|}
\hline
\textbf{No.} & \textbf{Category} & \textbf{Features} & \textbf{Explanation} & \textbf{\#} \\
\hline
1 & Intonation & Frequency & Fundamental frequency, related to the pitch of the voice. & 1 \\
  & & MFCCs & Mel Frequency Cepstral Coefficients, capture timbral aspects that are crucial for intonation. & 13 \\
\hline
2 & Volume & Energy & Measures the signal's loudness. & 1 \\
  & & Entropy of Energy & Indicates variation in loudness within a frame. & 1 \\
\hline
3 & Rhythm & Zero Crossing Rate & Reflects the number of times the waveform crosses zero, related to the frequency of the signal. & 1 \\
\hline
4 & Rate & Rate of Speech & Measures how fast words are spoken. & 1 \\
  & & Number of Syllables & Counts the syllables, indicating speech density and pace. & 1 \\
\hline
5 & Pause & Number of Pauses & Total pauses, reflecting speech interruptions and flow. & 1 \\
  & & Balance & Ratio of speaking to pausing, indicates rhythmic flow. & 1 \\
\hline
6 & Spectral& Spectral Centroid & Center of gravity, affects perceived pitch and sharpness. & 1 \\
  & & Spectral Spread & Measures the width of the spectrum, related to the sharpness of sound. & 1 \\
  & & Spectral Rolloff & The frequency below which 90\% of energy lies, indicates the shape. & 1 \\
  & & Spectral Flux & Measures the changes between frames, indicates rhythm changes. & 1 \\
  & & Spectral Entropy & Reflects the entropy of spectral distribution, a complexity measure. & 1 \\
\hline  
7 & Chroma & Chroma & A set of 12 coefficients each representing a semitone within an octave, used in harmony analysis. & 12 \\
\hline

8 & Duration & Speaking Duration & measure speaking time (excluding fillers and pause) & 1 \\
  & & Original Duration & measure speaking time (including fillers and pause) & 1 \\
\hline

\hline
\end{tabular}
\caption{Detailed categorization of speech features into relevant categories, with explanations and specific feature counts, tailored for comprehensive speech pattern analysis in clinical assessments such as autism.}
\label{table:speech_features}
\end{table}

Each category of features captures different characteristics of speech that are potentially altered in ASD:
- \textbf{Prosody features} such as pitch (fundamental frequency) variations and speech rate are directly related to the emotional and syntactical aspects of speech, which are often atypical in ASD.
- \textbf{Energy and Zero Crossing Rate} provide basic information about the speech amplitude and frequency, which are useful for detecting abnormalities in speech loudness and pitch changes.
- \textbf{Spectral and Chroma features} reflect the quality of sound and harmony in speech. These features are sophisticated and can detect subtleties in speech that are not apparent through simple auditory observation.
- \textbf{MFCCs and their deltas} offer a robust representation of speech based on the human auditory system's perception of the frequency scales, essential for identifying nuanced discrepancies in how individuals with ASD perceive and produce sounds.

By analyzing these features using machine learning models, we aim to identify patterns that are indicative of ASD, thereby assisting in the objective and efficient diagnosis of the disorder.


\subsection{Diagnosis and Analysis of ASD Based on Gradient Boosting}
\label{sec:classification_asd}

This subsection elaborates on the use of various Gradient Boosting \cite{natekin2013gradient,bentejac2021comparative,ke2017lightgbm} to classify subjects into two groups defined by their scores on the ADOS-2 Module 4: individuals diagnosed with ASD (A2 score $\geq$ 1) were labeled as positive, and those not diagnosed with ASD (A2 score = 0) were labeled as negative. This binary classification task aimed to automate the preliminary screening process for ASD by identifying speech patterns that were characteristic of ASD as opposed to other speech irregularities.

To achieve this, we employed Gradient Boosting classifiers and regressors to analyze speech patterns for the diagnosis and assessment of Autism Spectrum Disorder (ASD). To ensure the reproducibility and robustness of our findings, we meticulously configured the parameters of our models. Both the classifier and regressor were set up with 100 estimators, a learning rate of 0.1, and a maximum depth of 3, allowing for sufficient complexity while avoiding overfitting. We used a minimum of 2 samples required to split an internal node and a minimum of 1 sample required at a leaf node, ensuring that each tree in the ensemble contributed to the learning process without biasing the results toward outliers. The algorithm was applied to the dataset comprising all extracted features, as detailed in Table \ref{table:speech_features}. To robustly estimate model performance and generalizability, we implemented a 5-fold cross-validation approach. This method involved dividing the entire dataset into five distinct subsets, using each in turn for testing while training the model on the remaining four-fifths of the data. This cross-validation procedure helped minimize overfitting and bias, ensuring that our findings were reliable and representative of diverse samples. By detailing these methodological choices, we provide a comprehensive framework that allows for the transparent and replicable application of machine learning techniques in clinical research.

To evaluate the efficacy of the model in our study, we utilized a comprehensive set of performance metrics: Accuracy, which measures the overall rate of correct predictions; Precision, indicating the correctness among positive predictions; Recall, assessing the model's ability to identify actual positives; and F1-score, which balances precision and recall for a comprehensive performance measure. These metrics allowed us to effectively gauge the models' precision and reliability in classification tasks.

\subsection{Feature Importance Evaluation through Regression Analysis}
\label{sec:regression_feature_importance}

Building upon the classification models discussed previously, we further employed regression analyses to evaluate the importance of speech features in predicting ASD. This step is crucial for understanding which features most significantly predict ASD outcomes, thereby providing insights that could inform clinical assessments and interventions.

The significance of each feature was quantified based on the magnitude and statistical significance of their coefficients derived from the regression models. Features with larger absolute values of coefficients were identified as having more substantial impacts on the ASD diagnosis. The performance of each regression model was evaluated using several metrics to ensure accuracy and reliability of the findings:
\begin{itemize}
    \item \textbf{Mean Absolute Error} (MAE): Represents the average magnitude of the errors in a set of predictions, without considering their direction (i.e., over or under-predicting).
    \item \textbf{Mean Squared Error} (MSE): Similar to MAE but squares the differences to punish larger errors more severely, which can be particularly useful in avoiding large prediction errors in a clinical setting.
    \item \textbf{R-Squared} (\(R^2\)): Indicates the proportion of variance in the dependent variable that is predictable from the independent variables; a higher \(R^2\) value suggests a better fit of the model to the data.
\end{itemize}

\section{Results}

\subsection{Analysis of Speech Pattern Features}

This subsection details the correlation analysis performed on various speech features extracted from the ADOS interviews. The features analyzed include numerical measures such as the number of syllables, pauses, speech rate, articulation rate, durations, and chromatic features across 12 semitones.

The analysis involved calculating the Pearson correlation coefficients between all pairs of selected speech features. This statistical measure helps in identifying the degree and direction of linear relationships between the features.

\begin{figure}[t]
 \centering
 \includegraphics[width=1\linewidth]{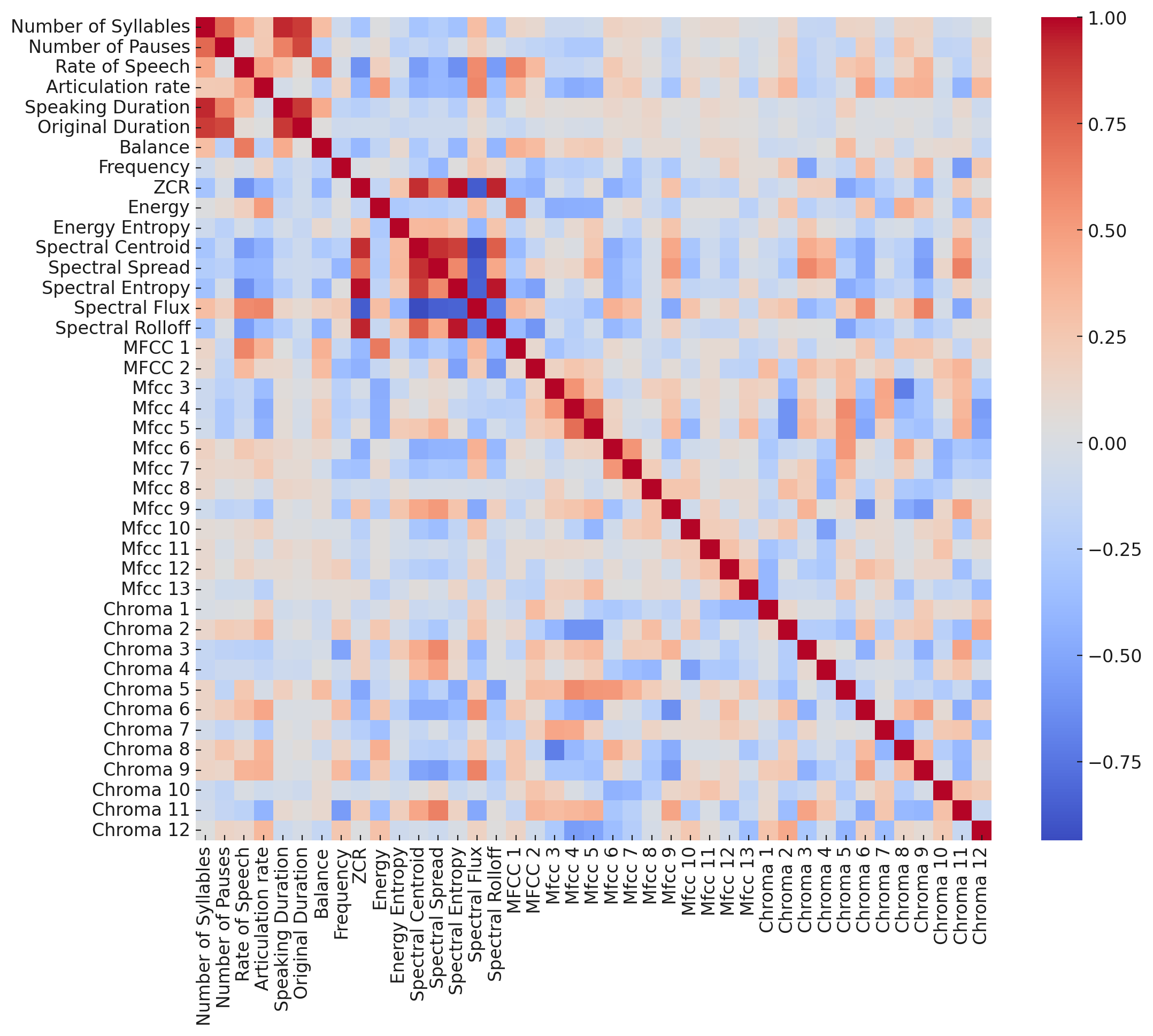}
 \caption{Correlation Matrix of Speech Features Extracted from ADOS Interviews.}
  \label{fig:1}
\end{figure}

The correlation matrix revealed several key insights:
\begin{itemize}
    \item \textbf{High Correlation Between Rate Measures}: The rate of speech and articulation rate showed a strong positive correlation, suggesting that as participants speak faster, they tend to articulate more syllables per unit time. This relationship underscores the interdependence of these rate-based features.
    \item \textbf{Duration and Balance}: There was a notable correlation between speaking duration and original duration, indicating that longer speeches tend to have proportionally longer pauses. The balance ratio, which compares speaking duration to the total duration, helped in further highlighting differences in speech fluency.
    \item \textbf{Chromatic Relationships}: The chroma features, representing tonal energy distribution across different pitches, displayed varied degrees of correlation. Adjacent chroma features (e.g., Chroma 1 with Chroma 2) often showed moderate to high correlations, reflecting harmonic relationships in speech tonality.
\end{itemize}

Together, understanding these correlations helps in feature selection, reducing dimensionality, and mitigating issues related to multicollinearity in predictive modeling.

\subsection{Autism Diagnosis based on Speech Patterns}

This subsection presents the outcomes of employing various machine learning algorithms to diagnose ASD based on extracted speech features. The models tested include Support Vector Machine (SVM) \cite{rahman2020review}, Random Forest \cite{chi2022classifying}, AdaBoost \cite{albahri2023early}  , Naive Bayes \cite{rasul2024evaluation} , Multi-Layer Perceptron (MLP) \cite{mohanta2022analysis}, and Gradient Boosting \cite{cho2019automatic}. The effectiveness of each model was assessed using metrics such as accuracy, precision, recall, and F1-score, which are crucial for evaluating classification performance in medical diagnostic applications.

The performance of each model, summarized in Table \ref{tab:features}, indicates that Gradient Boosting outperformed other models with the highest accuracy and F1-score of 87.75\% and 89.89\%, respectively. These metrics are particularly important in the medical field to ensure the diagnostic predictions' reliability and relevance.

\begin{table}[]
\centering
\begin{tabular}{@{}lllll@{}}
\toprule
Model      & Accuracy &   Precision & Recall & F1-Score  \\ \midrule
SVM \cite{rahman2020review}                 & 83.08\%       & 82.88\%     & 93.04\%         & 86.91\%  \\
Random Forest \cite{chi2022classifying}     & 87.01\%       &  83.27\%     & \textbf{98.16\%}         & 89.69\%  \\
AdaBoost \cite{albahri2023early}            & 86.64\%       & \textbf{84.93\%}     & 95.16\%         & 89.35\%  \\
Naive Bayes \cite{rasul2024evaluation}         & 84.36\%       & 81.83\%     & 95.51\%         & 87.85\%  \\
MLP \cite{mohanta2022analysis}                 & 82.15\%       & 84.29\%     & 90.15\%         & 86.09\%  \\ 
Gradient Boosting \cite{cho2019automatic} & \textbf{87.75\%}       & 84.73\%     & 96.96\%         & \textbf{89.89\%}  \\ \bottomrule
\end{tabular}
\caption{Comprehensive Speech Features Extracted for Analyzing ASD.}
\label{tab:features}
\end{table}

The best-performing model, Gradient Boosting, yielded a confusion matrix as follows:

\begin{figure}[h!]
\begin{center}
\includegraphics[width=12cm]{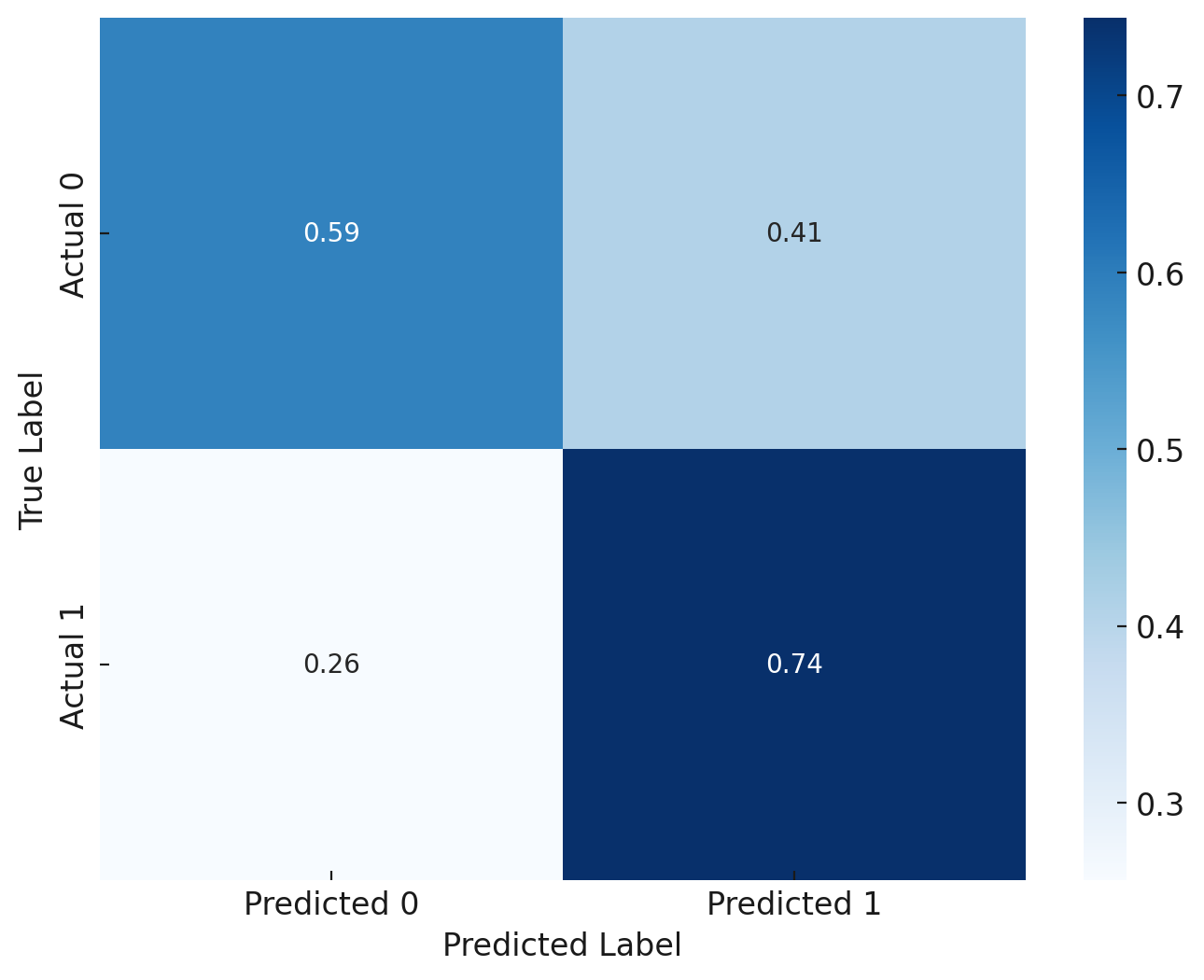}
\end{center}
\caption{Confusion matrix for Gradient Boosting.}\label{fig:2}
\end{figure}

This matrix reflects the model’s performance in classifying the ASD diagnoses. The high recall (96.96\%) of the Gradient Boosting model suggests it is highly capable of identifying most cases of ASD, which is critical to ensuring that individuals needing intervention are not overlooked. However, the presence of false positives as indicated by precision (84.73\%) suggests some individuals might be incorrectly diagnosed with ASD, underscoring the need for secondary confirmation of cases.

The results indicate that machine learning models, especially Gradient Boosting, can significantly enhance the accuracy and efficiency of ASD diagnostics based on speech patterns. Future work will focus on refining these models, possibly by integrating additional types of behavioral data, to further improve diagnostic accuracy and reduce false positives.

The use of advanced machine learning models to analyze speech patterns holds promise for improving the diagnosis of ASD. The results from this study emphasize the potential of these models to serve as effective tools in clinical settings, aiding in the early detection and accurate diagnosis of ASD. Further research is required to optimize these models for clinical use, ensuring they complement traditional diagnostic methods effectively.

\subsection{Analysis of Feature Importance}

This subsection discusses the application of regression models to determine the importance of various speech features in diagnosing ASD. The regression models not only predict ASD but also highlight which features are most indicative of the disorder, thus providing critical insights into the key aspects of speech that differentiate ASD cases from non-ASD cases.

Table \ref{tab:regression} summarizes the performance of various regression models applied to the speech features. These models were evaluated based on their \(R^2\) (Coefficient of Determination), Mean Absolute Error (MAE), and Mean Squared Error (MSE), which indicate the models' accuracy, error magnitude, and error variance, respectively.

\begin{table}[]
\centering
\begin{tabular}{@{}llll@{}}
\toprule
Model & \( R^2 \)    & MAE & MSE    \\ 
\midrule
Naive Bayes \cite{rasul2024evaluation} & 0.513  & 0.098 & 0.098     \\ 
MLP \cite{mohanta2022analysis} & 0.698  & 0.192 & 0.061                \\ 
AdaBoost \cite{albahri2023early}  & 0.774  & 0.185 & 0.045                 \\ 
SVM \cite{rahman2020review} & 0.860 & 0.124 & 0.028               \\
Random Forest \cite{chi2022classifying}  & 0.857 & 0.083 & 0.029               \\
Gradient Boosting \cite{cho2019automatic} & \textbf{0.881}  & \textbf{0.094} & \textbf{0.024}            \\

\bottomrule
\end{tabular}
\caption{Regression based on Speech Features Extracted for Analyzing ASD.} 
\label{tab:regression}
\end{table}

The \(R^2\) values indicate the proportion of variance in the ASD diagnosis that can be predicted from the speech features. Gradient Boosting provided the highest \(R^2\) value, suggesting it is the most effective model in terms of explaining the variance in ASD diagnoses from the speech features. Additionally, it achieved the lowest MAE and MSE, indicating high accuracy and reliability with minimal error in predictions.

\begin{figure}[t]
 \centering
 \includegraphics[width=1\linewidth]{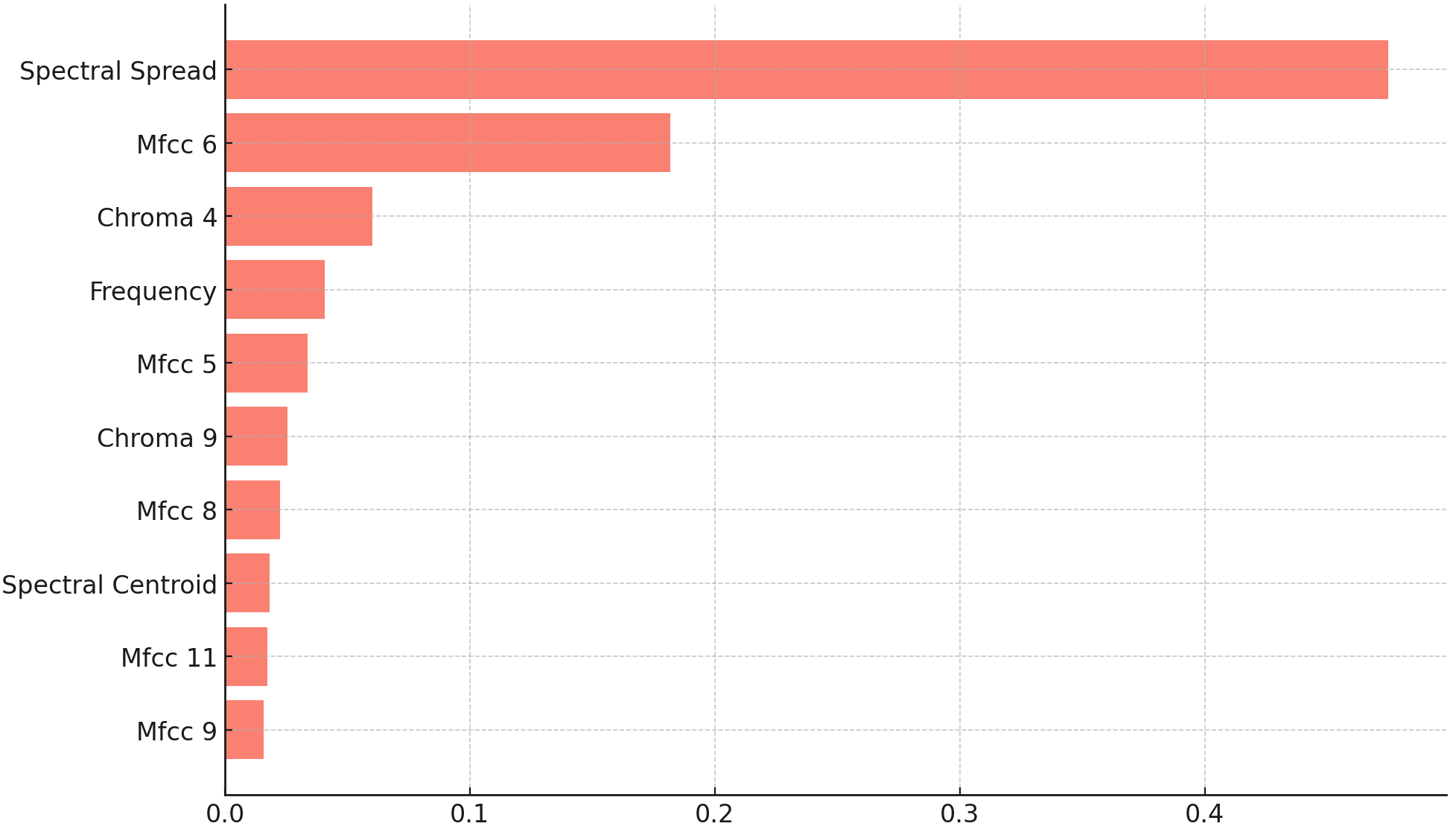}
 \caption{Top 10 Features in Regression Analysis Based on MDI Importance Scores.}
  \label{fig:mdi_ranking}
\end{figure}

The regression analysis also involved examining which speech features contributed most significantly to diagnosing ASD. This was assessed using the feature importance scores derived from the best-performing model gradient boosting, as visualized in Figure \ref{fig:mdi_ranking}.

This graph ranks the speech features according to their importance, determined by the magnitude of their coefficients in the model. Features with higher scores have a greater impact on the model's output and are, therefore, more critical for predicting ASD.

\section{Discussion}
This study explored the efficacy of various machine learning models in diagnosing ASD through speech patterns, highlighted the importance of specific speech features, and assessed the potential clinical applications of these findings. Here, we'd like to discuss the implications of our results, the limitations of the current study, and propose directions for future research.

Our analysis revealed that the Gradient Boosting model exhibited the highest accuracy and F1-score among the tested algorithms, suggesting its superiority in capturing complex, nonlinear relationships between features. This underscores the suitability of ensemble methods like Gradient Boosting \cite{bentejac2021comparative}, Random Forest \cite{breiman2001random}, and AdaBoost \cite{freund1997decision} for the nuanced task of ASD diagnosis through speech, given their capability to integrate multiple decision trees for improved prediction accuracy while mitigating overfitting. Furthermore, features such as speech rate, articulation rate, and specific chromatic aspects, particularly lower chroma features, emerged as critical in predicting ASD. These features likely capture nuances in speech dynamics and the characteristic of tonal variation of ASD speech patterns, such as atypical intonation and rhythm, which aligns with the literature suggesting distinct prosodic behaviors in individuals with ASD \cite{asghari2021distinctive}.

The potential application of non-invasive, speech-based markers for ASD diagnosis, as indicated by our findings, could significantly lower barriers to early screening and assessment \cite{pokorny2017earlier}. This holds particular relevance in settings with limited traditional diagnostic resources or when individuals are less responsive to conventional diagnostic procedures such as ADOS \cite{lord1999ados}. Moreover, understanding which speech features are most indicative of ASD can inform tailored interventions targeting specific communicative impairments. For instance, therapies designed to enhance speech fluidity and intonation could be beneficial for individuals exhibiting significant deviations in these areas \cite{vogindroukas2022language}.  Interventions that directly target speech prosody using established evidence-based practices for ASD may be most effective for increasing typical prosodic patterns during speech for persons with ASD \cite{holbrook2020speech}. 

While our models performed well on the dataset used, their generalizability to other populations, languages, and cultural contexts remains untested. Future studies should aim to validate these models in more diverse demographic settings to enhance their applicability \cite{hsiao2018autism}. Additionally, integrating speech-based diagnostics with other behavioral and neurological indicators may enhance the accuracy and reliability of ASD assessments. Exploring multimodal diagnostic frameworks that combine speech features with facial expression analysis and motor patterns could be fruitful \cite{han2022multimodal}. Furthermore, advancements in machine learning, including the development of more sophisticated neural network architectures, could refine the accuracy of speech-based ASD diagnosis. Deep learning models capable of processing sequential and temporal information more effectively may unveil more complex patterns indicative of ASD.

\section{Conclusion}

This study illustrated the potential of machine learning models, particularly Gradient Boosting, to enhance ASD diagnosis through speech pattern analysis, achieving high accuracy and demonstrating significant differentiation between ASD and non-ASD cases based on speech features alone. Key findings highlighted the importance of specific speech features such as the rate of speech and articulation rate, which are markedly indicative of ASD and capture essential speech dynamics like intonation and rhythm. These machine learning techniques, especially sophisticated ensemble methods, present non-invasive, efficient, and accessible means for early ASD screening, which is crucial for timely interventions that markedly improve developmental outcomes. Future research should aim to validate these models across diverse demographics and integrate more complex neural network architectures to handle the intricate patterns in speech data better. Overall, this research underscores the transformative potential of machine learning in clinical diagnostics, offering robust tools for early detection and opening new avenues for automated, non-intrusive, and comprehensive ASD screening on a global scale.



\section*{Conflict of Interest Statement}

The authors declare that the research was conducted in the absence of any commercial or financial relationships that could be construed as a potential conflict of interest.

\section*{Author Contributions}

CH works on all aspects of this work including data processing, experiments, result analysis, and paper writing. JT and WL participated in data processing and experiments. XY and LP helped with the result analysis. SW and XL are responsible for project monitoring, funding support, paper writing and revision.

\section*{Funding}
This project is partially supported by NSF HCC-2401748 and Prof. Xin Li's start-up funds from UAlbany.




\bibliographystyle{Frontiers-Harvard} 
\bibliography{test}


\section*{Figure captions}



\end{document}